\documentclass[conference]{IEEEtran}
\IEEEoverridecommandlockouts
\usepackage{cite}
\usepackage{amsmath,amssymb,amsfonts}
\usepackage{algorithmic}
\usepackage{graphicx}
\usepackage{textcomp}
\usepackage{subfig}
\usepackage{algorithm}
\usepackage{xcolor}
\usepackage{booktabs}
\usepackage{setspace}
\newcommand{\blue}[1]{\textcolor{black}{#1}}
\usepackage{siunitx}
\usepackage[numbers,sort&compress]{natbib}

\usepackage[draft=true]{hyperref}
\graphicspath{ {./figures/} }
\def\BibTeX{{\rm B\kern-.05em{\sc i\kern-.025emb}\kern-.08em}}

\begin{document}

%
\title{
\huge{Personalized Federated Learning-Driven Beamforming Optimization for Integrated Sensing and Communication Systems}
}

\author{
    \IEEEauthorblockN{Zhou Ni\IEEEauthorrefmark{1}, Sravan Reddy Chintareddy\IEEEauthorrefmark{1}, Peiyuan Guan\IEEEauthorrefmark{2}, Morteza Hashemi\IEEEauthorrefmark{1}}
\IEEEauthorblockA{\IEEEauthorrefmark{1}Department of Electrical Engineering and Computer Science, University of Kansas, Lawrence, USA, \\
\IEEEauthorrefmark{2} Department of Informatics, University of Oslo, Oslo, Norway}

}

\maketitle

\begin{abstract}
In this paper, we propose an Expectation-Maximization-based (EM) Personalized Federated Learning (PFL) framework for multi-objective optimization (MOO) in Integrated Sensing and Communication (ISAC) systems. In contrast to standard federated learning (FL) methods that handle all clients uniformly, the proposed approach enables each base station (BS) to adaptively determine its aggregation weight with the EM algorithm. \blue{Specifically, an EM posterior is computed at each BS to quantify the relative suitability between the global and each local model, based on the losses of models on their respective datasets.} The proposed method is especially valuable in scenarios with competing communication and sensing objectives, as it enables BSs to dynamically adapt to application-specific trade-offs. \blue{To assess the effectiveness of the proposed approach, we conduct simulation studies under both objective-wise homogeneous and heterogeneous conditions.} The results demonstrate that our approach outperforms existing PFL baselines, such as FedPer and pFedMe, achieving faster convergence and better multi-objective performance.
\end{abstract}


\section{Introduction}
Federated learning (FL) has emerged as a decentralized paradigm for privacy-preserving collaborative learning across devices. FL enables participants (devices or servers) to train shared models while maintaining data locally, thereby mitigating privacy risks and eliminating centralization requirements \cite{mcmahan2017communication}.  
Despite its benefits, FL encounters several inherent challenges from  
data heterogeneity across devices, which can cause performance variations among local models~\cite{li2020federated}.
Additionally, FL efficiency is also limited by communication overhead from continuous server-client updates~\cite{konevcny2016federated, chen2024optimizing}. 
To address these limitations, personalized federated learning (PFL) strategies have been developed to learn models aligned with individual clients' data characteristics. For instance, in \cite{deng2020adaptive}, the authors propose an adaptive algorithm that combines local and global models to address statistical heterogeneity. In \cite{ni2025pfedwn}, a server-free framework is proposed for device-to-device (D2D) wireless networks that incorporates channel-aware neighbor selection and weight assignment to handle non-IID and unbalanced data distributions while optimizing resource allocation and interference management.

Concurrently, ISAC has emerged to enhance both communication and sensing performance~\cite{zhou2022integrated}. 
\blue{ISAC addresses spectrum utilization and sensing capability requirements in diverse wireless systems, ranging from cellular networks and vehicular communications to autonomous platforms such as drones and self-driving vehicles.} By sharing spectrum resources and improving hardware-software interoperability, ISAC enables efficient network utilization and supports advanced applications in smart environments~\cite{wang2022integrated}. However, 
\blue{optimal ISAC operation is impeded by \textbf{(i)} cross-domain interference between sensing and communication, \textbf{(ii)} jointly coupled beamforming and resource allocation under power constraints, and \textbf{(iii)} strong inter-BS heterogeneity (non-IID data or objectives) that invalidates a single global model.} These issues are amplified in distributed environments lacking centralized coordination. Given these distributed challenges, FL presents a natural solution by enabling distributed decision-making and resource optimization in ISAC networks. This approach preserves data privacy while improving adaptability, thereby enhancing system resilience in dynamic environments.


Nevertheless, while FL offers a promising theoretical solution for ISAC, recent studies reveal suboptimal performance in realistic scenarios with unbalanced or heterogeneous data. For instance, in \cite{liu2022joint}, the authors demonstrate that non-IID data across devices can lead to divergent local model parameters and inconsistent updates, preventing the global model from reaching convergence. Similarly, in \cite{liang2024federated}, the authors investigate FL in an integrated sensing, communication, and computation (ISCC) framework and find that the standard FedAvg approach (with multiple local updates) suffers from increased data heterogeneity. These findings highlight a significant barrier to the practical deployment of FL in ISAC networks. This performance gap \blue{motivates} strategies that can effectively manage statistical heterogeneity while ensuring robustness in dynamic environments. More recently, \blue{\cite{jiang2025federated} proposes two FL frameworks for coordinated beamforming across BSs in multi-cell ISAC systems. While their solutions help mitigate inter-cell interference without sharing raw channel data, they primarily emphasize communication-centric coordination and do not address the personalized multi-objective trade-off between communication and sensing under heterogeneity.} To the best of our knowledge, this is the first work that applies PFL to ISAC networks for improving the system-level performance. Our main contributions in this paper are as follows:
\begin{itemize}
    \item We introduce a PFL framework specifically designed for ISAC networks, enabling effective joint optimization of radar sensing and communication objectives under heterogeneous local conditions.
    \item We \blue{develop an EM-based algorithm} that dynamically estimates personalized aggregation weights for each base station, improving model personalization in non-IID environments.
    \item Through comprehensive simulations, we demonstrate that our proposed EM-based PFL algorithm achieves improvements in joint communication and radar sensing performance, surpassing traditional FL by up to $17.81$\% and $16.79$\% \blue{of system weighted sum rate} in homogeneous and heterogeneous scenarios, respectively.
\end{itemize}

The rest of this paper is organized as follows. In Section \ref{sec:problem}, we present the system model and problem formulation. Section \ref{sec:solution} introduces the proposed EM-based PFL algorithm. Section \ref{sec:results} presents our numerical results, and Section \ref{sec:concludes} concludes the paper. 

\section{System Model and Problem Formulation}
\label{sec:problem}
\subsection{System Model}
\subsubsection{ISAC Communication Model}
\begin{figure}
    \centering
    \includegraphics[width=0.8\linewidth]{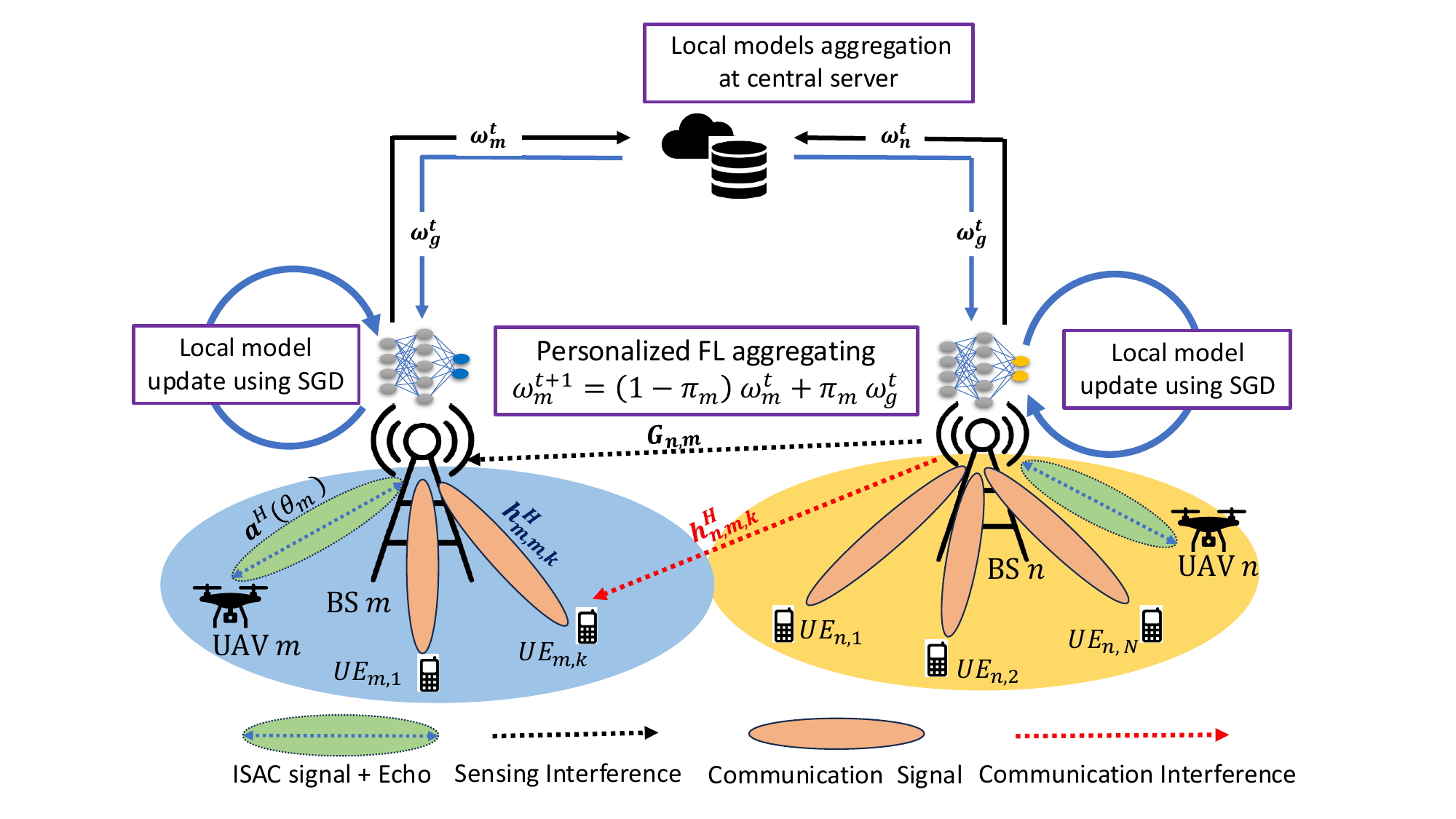}
    \caption{Proposed EM-based PFL multi-cell ISAC System. }
    \label{fig:jsca}
\end{figure}

We consider an ISAC system comprising multiple cells as shown in Fig.~\ref{fig:jsca}. Each cell consists of a BS, denoted by index $m \in \{1,2, ..., M\}$, that is equipped with $N_{T}$ antennas and simultaneously serves multiple communication users (UEs) while performing radar sensing of a \blue{local target (e.g., a UAV, vehicle, or infrastructure reflector)}. In each cell, we denote $k \in \{1,2, ..., K\}$ as the number of single-antenna communication users served by the $m$-th BS. For simplicity, we use the same index $m$ for both the BS and its corresponding radar target. 
Let $\textbf{X}_m = [\textbf{x}_{1},\textbf{x}_{2}, ..., \textbf{x}_{L}] \in \mathbb{C}^{N_{T} \times L}$ be the  signal matrix where $L$ is the communication frame length. Thus, we have the transmitted signal as:
\begin{equation}
    \textbf{X}_m = \textbf{W}_{m}\textbf{S}_m = \sum_{k = 1}^{K} \textbf{w}_{m,k}\textbf{s}_{m,k}^{H},
\end{equation}
where $\textbf{W}_{m} = [\textbf{w}_{m,1},\textbf{w}_{m,2},...,\textbf{w}_{m,K}]\in\mathbb{C}^{N_{T}\times K}$ is the designed beamforming matrix, and \blue{$\textbf{S}_m$ is the per-user symbol matrix whose $k$-th row is $\textbf{s}_{m,k} \in \mathbb{C}^{L\times 1}$ represents the data stream intended for the 
$k$-th UE in the $m$-th cell of frame length $L$ and normalized as $L^{-1}\|\mathbf{s}_{m,k}\|^2=1$.}

The received signal at the UE $k$ at cell $m$ can be written as follows:
\begin{equation}
\begin{aligned}
    \textbf{y}_{m,k}^{H} &= \textbf{h}_{m,m,k}^{H}\textbf{X}_{m} + \sum_{i \neq m}^{M}\textbf{h}_{i,m,k}^{H}\textbf{X}_{i} + \textbf{n}_{m,k}^{H}    \\
    &=\textbf{h}_{m,m,k}^{H}\textbf{w}_{m,k}\textbf{s}_{m,k}^{H} + \sum_{j\neq k}^{K}\textbf{h}_{m,m,k}^{H}\textbf{w}_{m,j}\textbf{s}_{m,j}^{H}  \\
    &+ \sum_{i \neq m}^{M}\sum_{j = 1}^{K}\textbf{h}_{i,m,k}^{H}\textbf{w}_{i,j}\textbf{s}_{i,j}^{H} + \textbf{n}_{m,k}^{H} ,
\end{aligned}
\end{equation}
where $\textbf{h}_{m,m,k}^{H} \in \mathbb{C}^{1 \times N_T}$ represents the communication channel vector between BS $m$ and its associated user $k$. $\textbf{h}_{i,m,k}^{H} \in \mathbb{C}^{1 \times N_T}$ denotes the interfering channel vector from BS $i$ (where $i \neq m$) to user $k$ in cell $m$. $\textbf{n}_{m,k}^{H} \in \mathbb{C}^{1\times L}$ is the additive white Gaussian noise (AWGN), \blue{such that} $ \textbf{n}_{m,k} \sim \mathcal{CN}(0, \sigma_c^2 I) \in \mathbb{C}^{L \times 1}$. Therefore, the SINR of the communication model at $k$-th UE in cell $m$ can then be written as:
{\small
\begin{equation}
    \zeta_{m,k}^{(c)} = \frac{ |\textbf{h}_{m,m,k}^H \textbf{w}_{m,k}|^2}{\sum_{j \neq k}  |\textbf{h}_{m,m,k}^H \textbf{w}_{m,j}|^2 + \sum_{i \neq m} \sum_{j=1}^K |\textbf{h}_{i,m,k}^H \textbf{w}_{i,j}|^2 + \sigma_c^2}.
\end{equation}
}

Thus, the achievable sum communication rate for BS $m$ is given by:
\begin{equation}\label{eq.4}
    R_m^{(c)} = \sum_{k = 1}^{K} \log_2{(1 + \zeta_{m}^{(c)})}.
\end{equation}

\subsubsection{ISAC Radar Sensing Model}

We model each radar target as a point target located in the coverage area of the BS, and each BS senses exactly one radar target at a given time. Specifically, by transmitting the ISAC signal matrix $\textbf{X}_m$, the $m$-th BS observes the reflected echo signal matrix at its receiver as:
\begin{equation}
    \textbf{Y}_m^{(s)} = \textbf{G}_m \textbf{X}_m + \sum_{n \neq m}^{M}\textbf{G}_{n,m}\textbf{X}_n + \textbf{G}_{m,m}\textbf{X}_{m} +  \textbf{N}_m^{(s)}  \in \mathbb{C}^{N_R \times L},
\end{equation}
where $\textbf{Y}_m^{(s)} \in \mathbb{C}^{N_R \times L}$ is the received radar echo signal matrix at BS $m$, and $\textbf{G}_m$ the target reflection channel matrix for the radar target associated with BS $m$. Additionally, $\textbf{G}_{n,m} \in \mathbb{C}^{N_R \times L}$ is the inter-cell radar interference channel matrix from the $n$-th BS ($n \neq m$) to the radar receiver of BS $m$ and $\textbf{G}_{m,m} \in \mathbb{C}^{N_R \times L}$ represents the self-interference channel.

In particular, the radar target response matrix $\textbf{G}_m$ can be modeled as follows:
\begin{equation}
    \textbf{G}_m = \beta_m \textbf{a}(\theta_m)\textbf{b}^{H}(\theta_m),
\end{equation}
where $\beta_m$ is the radar cross section (RCS) coefficient of the radar target in the $m$-th cell, which is modeled as a complex Gaussian random variable, i.e., $\beta_m \sim \mathcal{CN}(0, \alpha_s)$, where $\alpha_s$ is the mean reflection strength of the radar target. In addition, $\textbf{a}(\theta_m)$ and $\textbf{b}(\theta_m)$ are the receive and transmit array steering vectors at BS $m$, respectively.

\blue{To explicitly account for the network-level radar sensing interference, the received signal model at the $m$-th BS, including the echoes from neighboring cells, the echo signal reflected by the target can be written as:}
\begin{equation}
\begin{aligned}
    \textbf{y}_m^{(s)}(t) &= \sum_{n \neq m} \textbf{G}_{n,m} \textbf{W}_n \textbf{s}_n^H(t - \tau_{n,m}) + \textbf{G}_{m,m} \textbf{W}_m \textbf{s}_m^H(t) \\ &+ \alpha_s \textbf{a}(\theta_m) \textbf{b}^H(\theta_m)\textbf{W}_m \textbf{s}_m^H(t - 2\tau_m) + \textbf{n}_m^{(s)}(t),
\end{aligned}
\end{equation}
where $\tau_{n,m}$ denotes the propagation delay between BS $n$ and BS $m$, and $\tau_{m}$ is the propagation delay associated with the sensing target in the $m$-th cell.

For radar signal processing at each BS, we employ the maximum ratio combining (MRC) beamforming~\cite{jiang2025federated}, where the $m$-th BS applies the beamformer $\textbf{v}_{m}^{H} = \textbf{a}^{H}(\theta_m) \ \in \mathbb{C}^{1\times N_R}$ to process the received radar echo, resulting in:
\begin{equation}
    \hat{\textbf{y}}_{m}^{(s)} = \textbf{v}_m^{H}\textbf{y}_m^{(s)}.
\end{equation}

Then, the radar sensing SINR at the $m$-th BS is:
\begin{equation}
    \zeta_m^{(s)} = N_{R}\frac{\sum_{k=1}^{K}|\textbf{v}_m^H \textbf{G}_m \textbf{W}_m|^2}{\sum_{n \neq m}^{M} \sum_{j =1}^{K}|\textbf{v}_m^H \textbf{G}_{n,m} \textbf{W}_{n,j}|^2 + \sigma_s^2},
\end{equation}
where $\sigma_s^2$ denotes the radar receiver noise power. Hence, the achievable sum radar information rate of BS $m$ is defined as:
\begin{equation}\label{eq.10}
    R_m^{(s)} = \log_2{(1 + \zeta_{m}^{(s)})}.
\end{equation}

\subsection{Problem Formulation}

In ISAC networks, optimizing conflicting objectives, such as maximizing communication rate (Eq.\eqref{eq.4}) and radar sensing accuracy (Eq.\eqref{eq.10}), forms a classic multi-objective optimization (MOO) problem~\cite{xu2025psmgd}, which yields a set of Pareto optimal solutions rather than a single optimum solution. To address this, we apply a scalarization approach by introducing a personalization hyperparameter $\rho_m \in [0,1]$ at each BS, converting the MOO into a single-objective problem. By fixing $\rho_m$ before optimization, each BS selects a desired trade-off point on its Pareto frontier based on local priorities, enabling consistent beamforming adapted to its specific needs.  Specifically, we formulate the optimization problem as maximizing the weighted sum of communication and radar sensing performance across all $M$ BSs:
\begin{equation}\label{eq.11}
\begin{aligned}
&\max_{\{\textbf{W}_m\}_{m=1}^{M}} \sum_{m = 1}^{M} [\rho_{m}R_{m}^{(c)} + (1-\rho_{m})R_{m}^{(s)}] \\
&\textrm{s.t.}  \quad   ||\textbf{W}_{m}||_{F}^{2} \leq P_T, \forall m,
\end{aligned}
\end{equation}
where $\textbf{W}_m \in \mathbb{C}^{N_T \times K_m}$ is the beamforming matrix of the $m$-th BS, $N_T$ is the number of transmit antennas, $K_m$ denotes the number of communication users served by the $m$-th BS, and $P_T$ represents the transmit power budget constraint at each BS. The constraint ensures that the transmit power at each BS is limited by a predefined power budget $P_T$. 

\blue{However, solving the optimization problem in Eq.~\eqref{eq.11} is generally challenging due to non-convex, interference-coupled rate objectives under per-BS power constraints and strong non-IID heterogeneity, which makes a single global model suboptimal. Thus, we propose the PFL framework described previously to obtain efficient and adaptive solutions, allowing each BS to optimally balance its dual functionality objectives based on local data characteristics and network environments.}

\section{Proposed Solution}
\label{sec:solution}
In this section, we propose an EM-based PFL solution to effectively solve the optimization problem formulated in the previous section.

\subsection{Expectation-Maximization based PFL}

We leverage the EM algorithm at each BS to determine an adaptive aggregation weight $\pi_m$. While each BS $m$ independently selects its personalization hyperparameter $\rho_m$, reflecting the trade-off between communication and radar sensing, the EM algorithm further personalizes the learning process by adaptively estimating how much each BS should learn from the global model. 

Specifically, the adaptive aggregation weight $\pi_m$ is determined locally at each BS $m$, based on how well the global model fits each BS's local data distribution. The EM algorithm accomplishes this adaptive determination through two iterative steps at each communication round:

\begin{itemize}
\item \textbf{E-STEP:} \blue{Each BS computes a posterior probability to find whether a local sample $b\in D_m$ is better explained by the \emph{global} model $\omega_g$ or the \emph{local} model $\omega_{l}$.} Specifically, for the BS $m$, the posterior probability is computed as: 
\begin{equation}
    \lambda_{m}^{(t+1)} = \frac{\exp{[-\kappa l(\omega_{g}^{(t)}| D_m)]}}{\exp{[-\kappa l(\omega_{g}^{(t)}| D_m)]} + \exp{[-\kappa l(\omega_{l}^{(t)}| D_m)]}},
    \label{eq:E-step}
\end{equation}
where $ l(\cdot| D_m)$ denotes the loss function evaluated on the local dataset $D_m$ using the global model parameters $\omega_{g}^{(t)}$ and $\kappa > 0 $ is the temperature parameter. 

\item \textbf{M-STEP:} The adaptive aggregation weight $\pi_m^{(t+1)}$ is updated based on the posterior probability from the E-step as follows:
\begin{equation}
    \pi_{m}^{(t+1)} = \frac{1}{B}\sum_{b}^{B}\lambda_{m,b}^{(t+1)}, 
    \label{eq:M-step-1}
\end{equation}
where $B$ is the batch size.
\end{itemize}

In our setting, each model lies along a Pareto front representing different trade-offs between communication and sensing performance. The posterior $\lambda_m$ serves as a responsibility estimator, assigning higher weight to the model that better aligns with the BS’s implicit trade-off preference, without requiring explicit tuning.



\subsection{Proposed PFL model}
We adopt a PFL approach to address heterogeneous channel conditions across BSs in the ISAC network. Specifically, we assume that each BS $m$ holds a local dataset consisting of communication and radar channel samples, denoted as $D_m = \{\textbf{H}_m, \textbf{G}_m \}$.


The proposed PFL framework allows each BS to generate personalized model parameters by adaptively integrating information from its own local model and global model updates. That is, at the $t$-th global communication round, the local model update for BS $m$ is computed as follows:
\begin{equation}
    \omega_{m}^{(t+1)} = (1 - \pi_m) \omega_{m}^{(t)} + \pi_m\omega_{g}^{(t)},
\end{equation}
where $\omega_{m}^{(t)}$ denotes the personalized model parameters of BS $m$ at the $t$-th communication round, and $\omega_{g}^{(t)}$ represents the aggregated global model obtained by the central server at the same round. 

Additionally, the personalized model parameters $\omega_{m}^{(t)}$ at each BS $m$ are updated locally using gradient descent. Specifically, within the $t$-th global round, the $j$-th local gradient descent iteration at BS $m$ is given by:
\begin{equation}
    \omega^{(t, j+1)}_{m} = \omega^{(t, j)}_{m} - \eta \nabla f_{m}(\omega_{m}^{(t, j)}), j \in [0,...,E-1],
\end{equation}
where $\eta$ denotes the local learning rate, $E$ is the total number of local epochs, and $f_{m}(\cdot)$ represents the local loss function of BS $m$. After $E$ local iterations at BS $m$, the resulting locally trained model is taken as the updated personalized model to be aggregated at the next global round, i.e., $\omega^{(t)}_{m} \triangleq \omega^{(t, E)}_{m}$.

\subsection{PFL-based DNN Training Framework}
To obtain the optimal beamforming matrices for each BS, \blue{we implement a DNN training framework within our EM-based PFL algorithm.} The personalized unsupervised loss at each BS $m$ in a given communication round can be defined as:

\begin{equation}
    l_m(\textbf{W}_m) = -\rho_m R_m^{(c)} - (1-\rho_m) R_m^{(s)}.
\end{equation}

Then, the overall personalized training objective at BS $m$ becomes:
\begin{equation}
    \min_{\textbf{W}_m} l_m(\textbf{W}_m), \quad \textrm{s.t.}  \quad   ||\textbf{W}_{m}||_{F}^{2} \leq P_T.
\end{equation}

 Therefore, the global personalized objective can be expressed as a summation of the personalized objectives across all BSs, effectively aggregating local preferences into a cohesive global optimization problem:
\begin{equation}
    \min_{\{\textbf{W}_m\}_{m=1}^{M}} \sum_{m=1}^{M}l_m(\textbf{W}_m), \quad \textrm{s.t.}  \quad   ||\textbf{W}_{m}||_{F}^{2} \leq P_T, \forall m.
\end{equation}

\section{Evaluation Metrics}
\label{sec:results}

\subsection{Datasets and Settings}
\label{subsec:ds}
\blue{In this section, we present the datasets, simulation settings, and evaluation metrics used to assess the performance of the proposed EM-based PFL framework.}

\label{subsubsec:dg}
\subsubsection{Training Data Generation} 
\begin{figure}[t!]
\centering  
\includegraphics[width=0.8\linewidth]{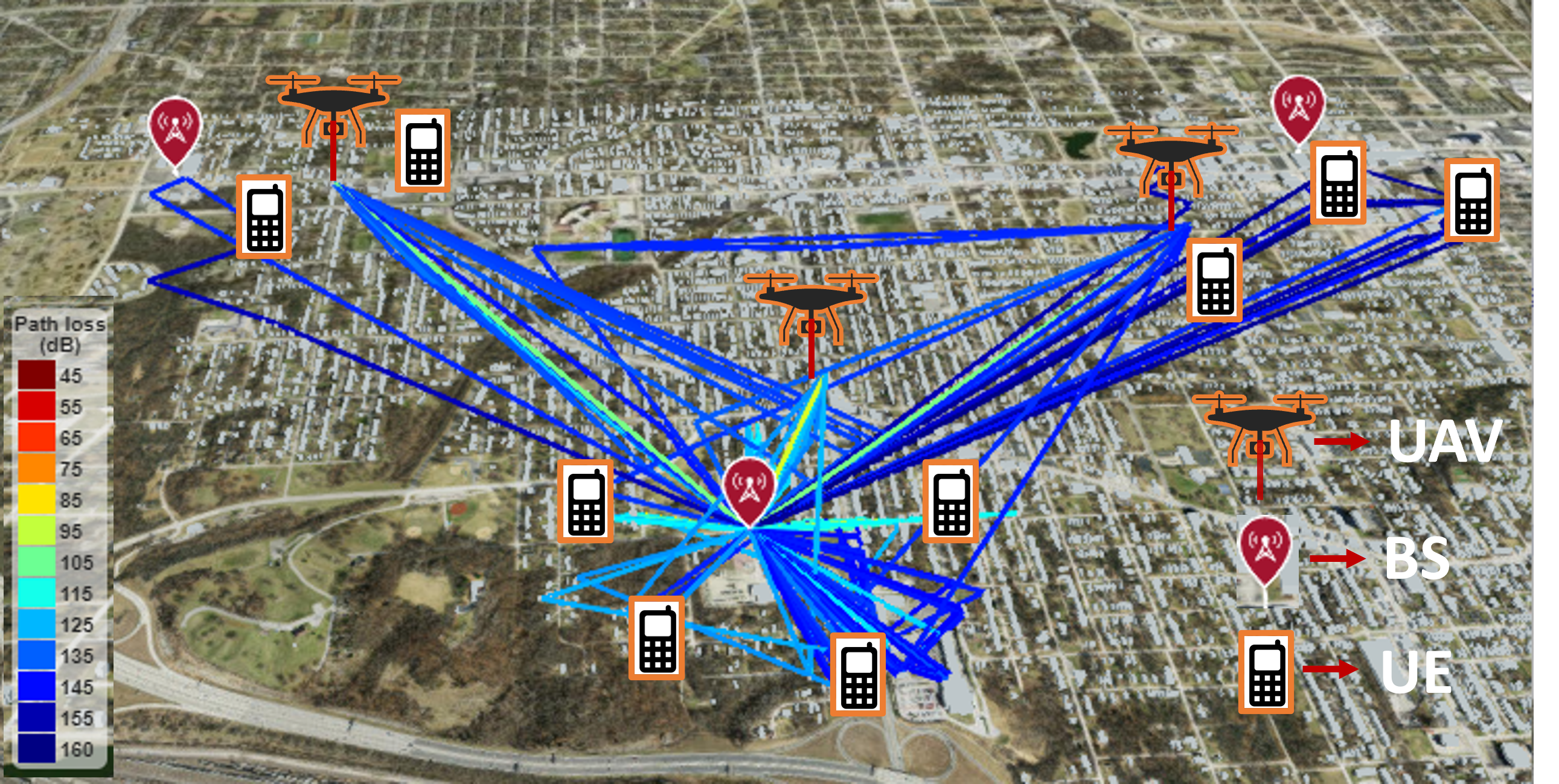}
    \caption{Ray-tracing simulation setup used for dataset generation. The plot illustrates the received signal paths at all the receivers transmitted from one base station.}    
    \label{fig:simsetup}
\end{figure}

As shown in Fig.~\ref{fig:simsetup}, we consider a multi-cell ISAC scenario with three neighboring cells, each served by a BS located at the cell center. Additionally, each BS is equipped with $N_T = 8$ transmit and $N_R = 8$ receive antennas, serving $2$, $3$, and $4$ communication users in the respective cells, while simultaneously sensing a radar target (modeled as an UAV). Users are randomly distributed within their respective cells, and radar targets are placed within an angular range of $[-\pi/2, \pi/2]$ relative to each BS. Both communication and sensing inter-cell interference channels follow Rician fading with a factor of $3$ to capture realistic multipath propagation.

To emulate realistic signal propagation, we adopt a ray-tracing channel model rather than relying on simplified stochastic models like Rayleigh or Rician fading alone~\cite{chintareddy2025federated}. We simulate a $3~\text{km} \times 3~\text{km}$ semi-urban area populated with buildings and vegetation, using OpenStreetMap~\cite{OSM} for map data and CellMapper~\cite{cellmap} for BS locations. MATLAB’s ray tracer is used to generate the wireless channels between each BS and the communication users and radar targets, providing a realistic propagation environment for multi-cell ISAC evaluation.

\subsubsection{DNN Architecture and Training Setup} \label{subsubsec:dnn_training}

To optimize the beamforming matrices, we leverage an ISAC-DNN architecture to optimize the beamforming matrices using federated datasets derived from MATLAB ray-traced channel tensors. For each BS \( m \), we extract a local dataset consisting of \( D_m = 20{,}000 \) channel samples, denoted as \( \{ \mathbf{H}_m^{(i)}, \mathbf{G}_m^{(i)} \}_{i=1}^{D_m} \). Here, \(\mathbf{H}_m^{(i)} \in \mathbb{R}^{N_T \times K_m \times 2}\) represents the stacked real–imaginary channel matrices for the \( K_m \) communication users, and \(\mathbf{G}_m^{(i)} \in \mathbb{R}^{N_T \times 1 \times 2}\) denotes the corresponding sensing channel tensors for the single UAV radar target within cell \( m \). To handle heterogeneity in user counts across cells (\(K_1 = 2\), \(K_2 = 3\), \(K_3 = 4\)), we perform zero-padding along the user dimension of \(\mathbf{H}_m^{(i)}\), aligning all channel tensors to a uniform user dimension of \( \max_m K_m = 4 \).

The designed ISAC-DNN comprises two parallel input branches, each employing a fully-connected layer with \(256\) neurons to independently process the flattened input communication and sensing channel samples. 
The final fully-connected layer outputs a real-valued vector of dimension \blue{$\mathbb{R}^{( N_T, \max_m K_m, 2)}$}, which is then reshaped and combined into complex-valued beamforming matrices \(\mathbf{W}_m \in \mathbb{C}^{N_T \times K_m} \). During inference, these matrices are normalized to ensure compliance with the transmit power constraint \(P_T\).

To facilitate personalized and robust model training, we utilize the proposed PFL approach over $100$ global communication rounds. Within each round, individual BSs perform local training for $5$ epochs, updating model parameters via the Adam optimizer with a learning rate of $10^{-4}$ and a batch size of $64$.

\begin{figure}[t]%
    \centering
    \subfloat[\centering EM-based weight assignment under homogeneous case.\label{subfig:homo}]{{\includegraphics[width=0.8\linewidth]{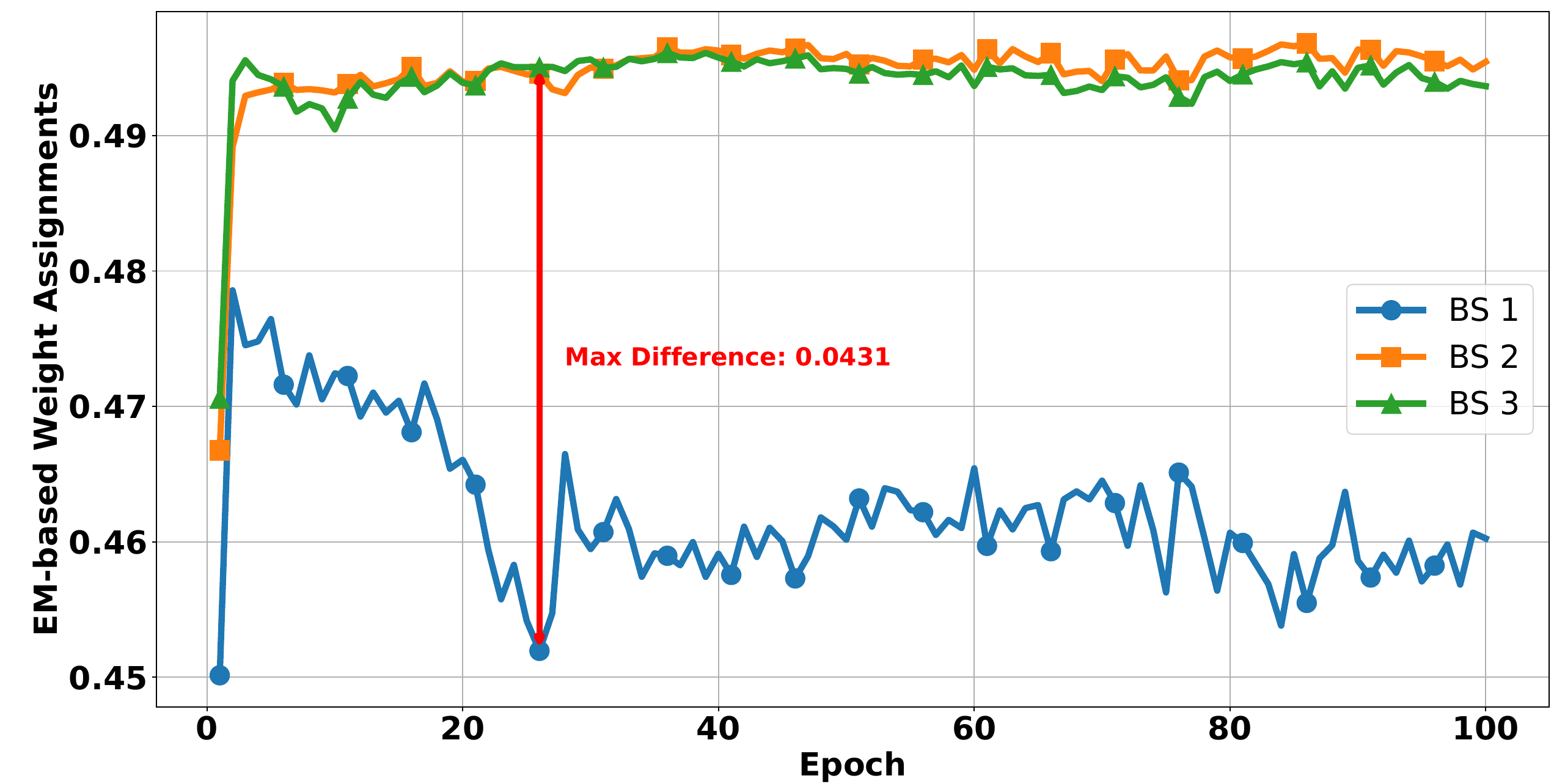} }}%
     \qquad
    \subfloat[\centering EM-based weight assignment under heterogenous case.\label{subfig:hete}]{{\includegraphics[width=0.8\linewidth]{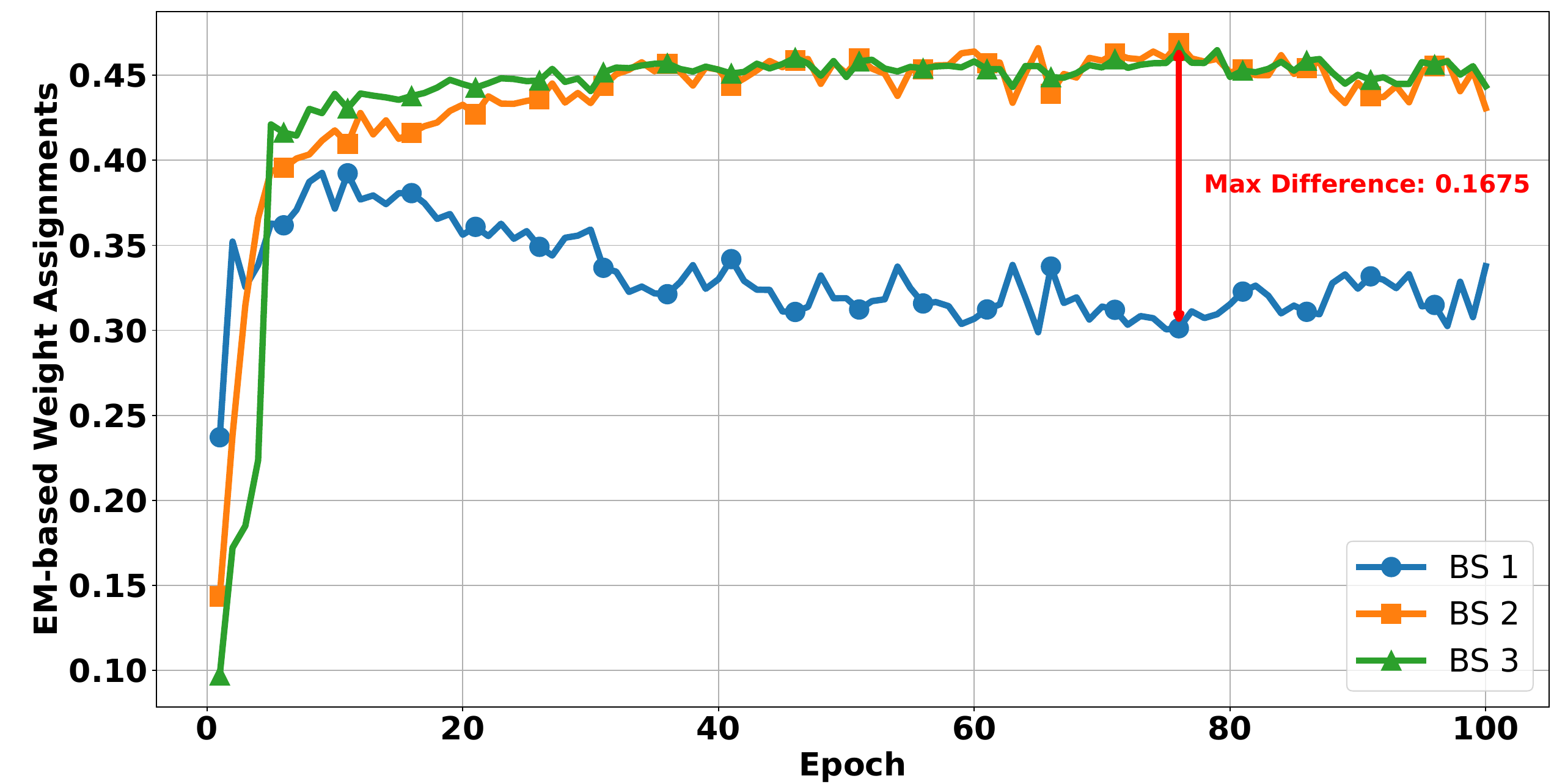} }}
    \caption{\small Proposed EM-based weight assignment methodology under two different cases.}%
    \label{fig:hete_weights}%
\end{figure}

\subsection{Evaluation Results}
\label{subsec:Eva}

We evaluate the effectiveness of our proposed PFL algorithm in comparison with four baseline methods: traditional FL~\cite{mcmahan2017communication}, pFedMe~\cite{t2020personalized}, FedPer~\cite{arivazhagan2019federated}, and fixed-weight PFL, which assigns fixed weights instead of dynamically computed EM weights. \blue{We evaluate each algorithm under both homogeneous and heterogeneous conditions, considering variations in learning objectives across BSs.} Specifically, in the homogeneous scenario, each BS has an identical personalization factor, $\rho_m = 0.5$, while in the heterogeneous scenario, the personalization factors are set differently across BSs, with $\rho_m$ values of $0.2, 0.6$, and $0.8$ for BS1, BS2, and BS3, respectively.


\subsubsection{EM-based Weight Assignment Efficiency}
First, the effectiveness of our proposed EM-based weight assignment method is demonstrated in Figures~\ref{fig:hete_weights}. In the homogeneous scenario (Fig.~\ref{subfig:homo}), $\pi_m$ converges closely across all BSs, with a smaller maximum difference of approximately $0.0431$. \blue{Even in this uniform scenario, our EM-based approach detects slight differences in local conditions. For example, BSs serving different numbers of communication UEs experience small variations in loss values, which the EM updates translate into marginal weight adjustments. These minor but systematic adaptations lead to improved convergence stability and overall system performance.} Conversely, in the heterogeneous scenario (Fig.~\ref{subfig:hete}), $\pi_m$ dynamically adapts to each BS’s unique conditions, resulting in significant and distinct variations across BSs, with a maximum observed difference of approximately $0.1675$ between BS1 and the others. This indicates that our method effectively captures and accommodates local heterogeneity based on the different objectives of each BS.

\subsubsection{Proposed EM-based PFL Performance Evaluation}
\begin{figure}[t] \centering
\includegraphics[width=0.8\linewidth]{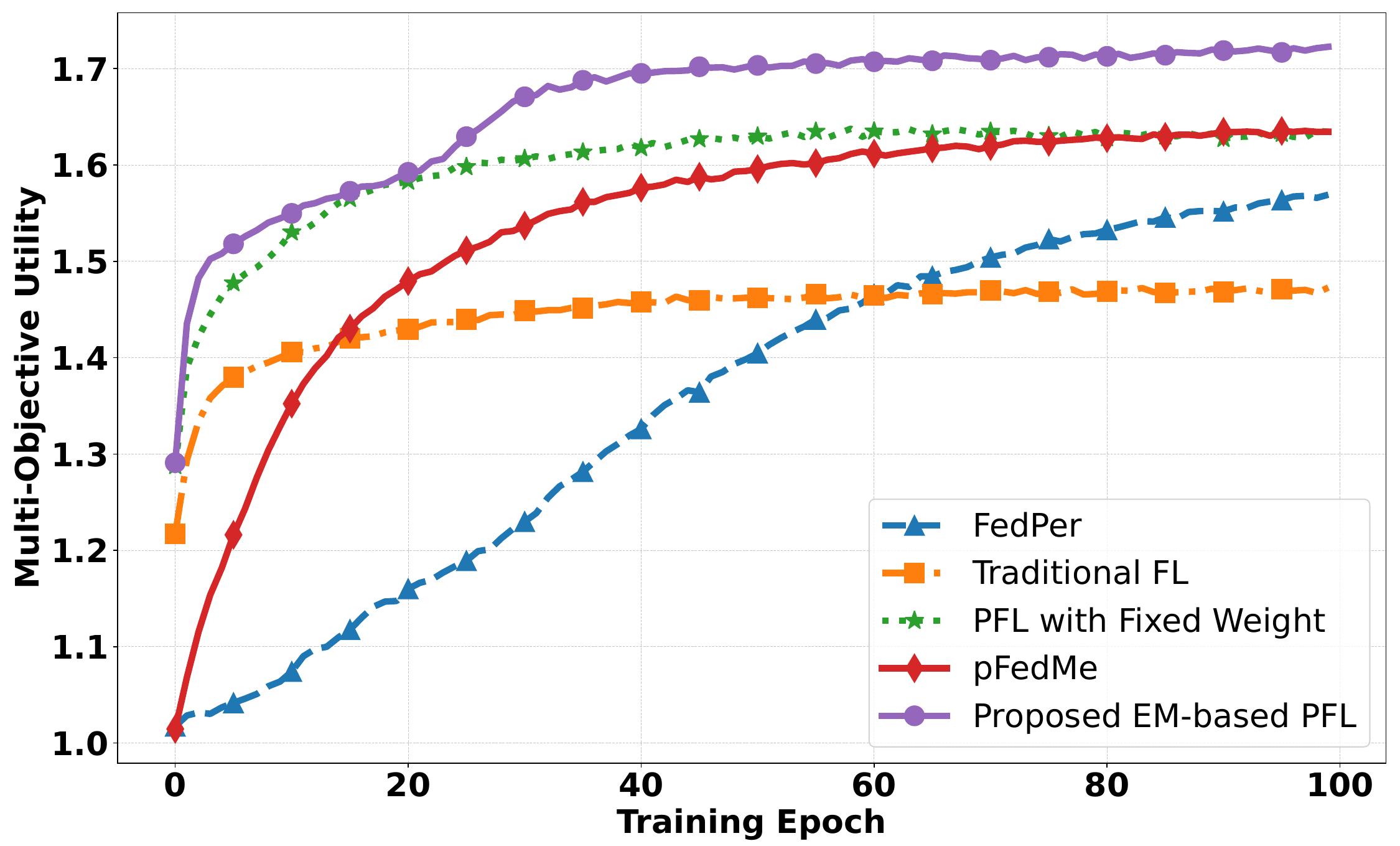} \caption{Multi-Objective Utility: Homogeneous Case.} \label{fig:homo_sum_rate} \end{figure}

\textbf{Homogeneous Scenario Analysis.} As presented in Fig.~\ref{fig:homo_sum_rate}, we observe that all algorithms rapidly converge within the first $20$ epochs except FedPer. Our proposed EM-weighted PFL algorithm achieves the highest system multi-objective utility, stabilizing at approximately $1.7$, and \blue{demonstrates the performance improvement over the baselines.} Fixed-weighted PFL follows closely, achieving a similar yet consistently lower performance, emphasizing the benefit of adaptive weight assignment. pFedMe exhibits moderate performance improvement compared to traditional FL, reflecting its advantage from personalization, yet it converges more slowly and achieves a significantly lower sum rate than our proposed method. Specifically, our algorithm achieves the highest utility of around $1.72$ \blue{in normalized multi-objective utility}, yielding improvements of $17.81$\% over traditional FL, $10.26$\% over FedPer, $5.52$\% over pFedMe, and $4.88$\% over fixed-weight PFL.

\begin{figure}[t] \centering
\includegraphics[width=0.8\linewidth,]{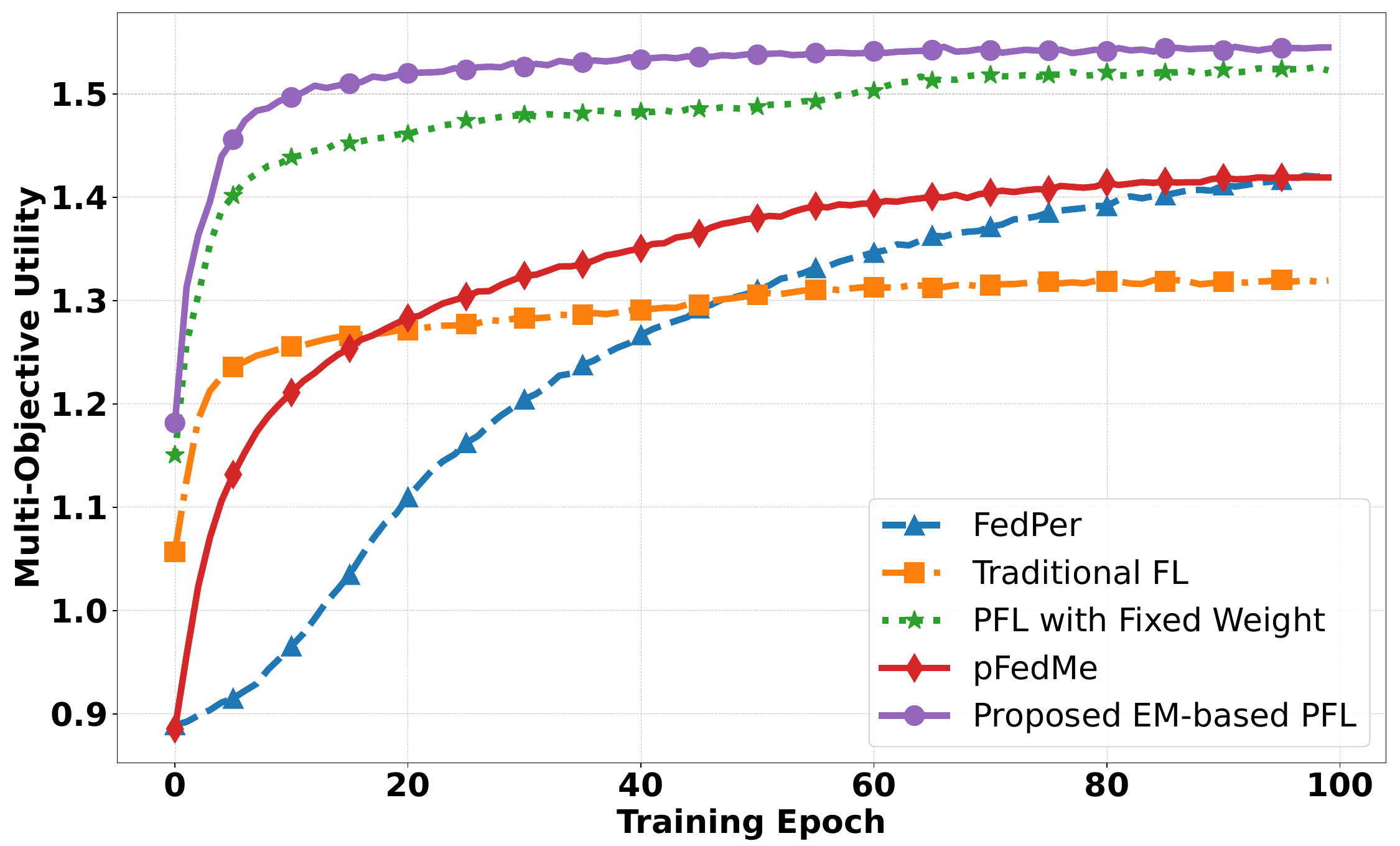} \caption{Multi-Objective Utility: Heterogeneous Case.} \label{fig:hete_sum_rate} \end{figure}

\textbf{Heterogeneous Scenario Analysis.} In the heterogeneous scenario shown in Fig.~\ref{fig:hete_sum_rate}, our proposed PFL algorithm again outperforms by rapidly converging and maintaining the highest sum rate of approximately $1.53$ \blue{in normalized multi-objective utility}, representing a $16.79$\% improvement over traditional FL, $8.51$\% over pFedMe, and $7.75$\% over FedPer. Fixed-weighted PFL achieves near-optimal results, though slightly lower than our dynamically EM-weighted approach. This highlights the necessity and effectiveness of adaptive weight computation based on local data distribution. pFedMe consistently outperforms traditional FL in this scenario as well, yet it remains inferior to both PFL variants in terms of final sum rate achieved and convergence speed.

In addition, it can be observed from Fig.~\ref{fig:homo_sum_rate} and~\ref{fig:hete_sum_rate} that our proposed PFL method starts with notably higher initial sum rates compared to traditional FL and other PFL algorithms. This behavior results from our algorithms employing immediate personalization and dynamic aggregation strategies from the very first epoch. Different from traditional FL and methods like pFedMe, which begin training from a purely global initialization with limited immediate adaptation to local data, our proposed method achieves a favorable local-global balance right from the initial aggregation. 

\begin{figure}[t]%
    \centering
    \subfloat[\centering Same number of UEs in each BS under homogeneous case.\label{subfig:homo_same}]{{\includegraphics[width=0.81\linewidth]{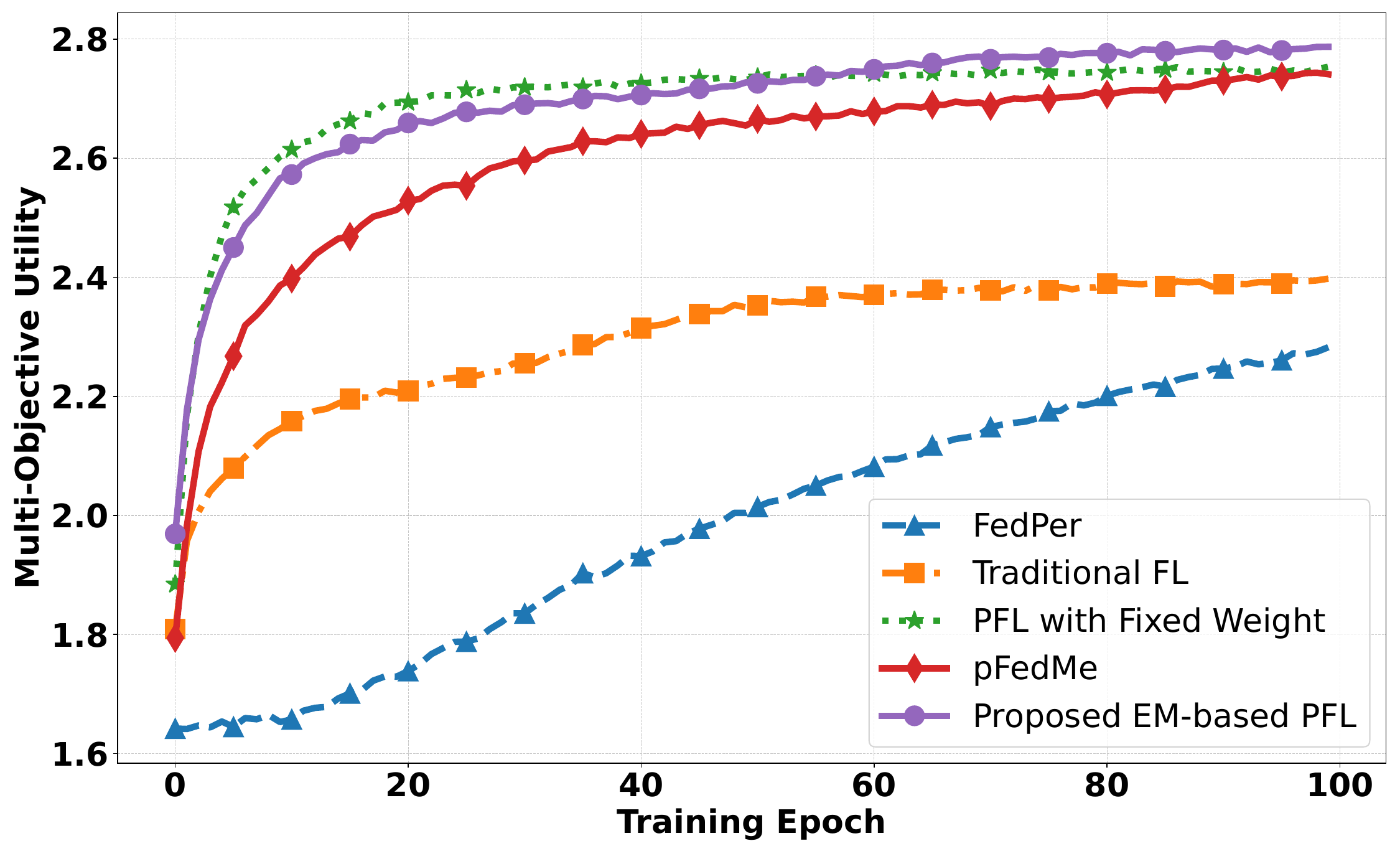} }}%
    \qquad
    \subfloat[\centering Same number of UEs in each BS under heterogenous case.\label{subfig:hete_same}]{{\includegraphics[width=0.8\linewidth]{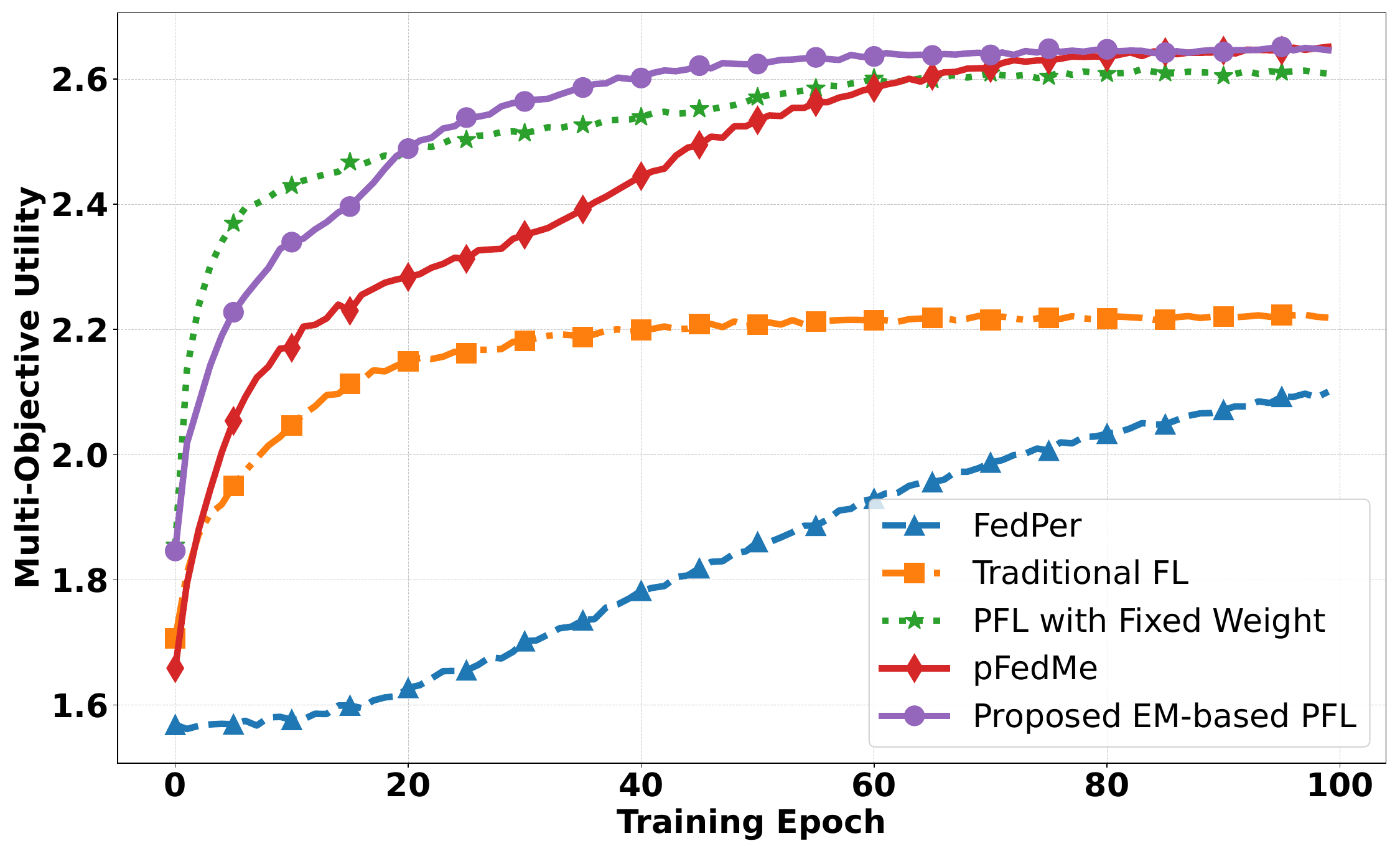} }}
    \caption{\small EM-based PFL performance under the condition where each BS serves the same number of UEs, shown for (a) the homogeneous case and (b) the heterogeneous case.}%
    \label{fig:same_UE}%
\end{figure}

To isolate the effect of BS heterogeneity from UE count variability, we conduct a controlled evaluation where each BS serves exactly two UEs. This uniform configuration removes the influence of UE number diversity, enabling a clearer comparison of PFL strategies under both homogeneous and heterogeneous objective conditions. As shown in Fig.~\ref{subfig:homo_same} and \ref{subfig:hete_same}, our EM-based PFL method consistently outperforms all baselines. Notably, our EM-based mechanism yields better performance than its fixed-weight PFL, indicating that even with equal UE counts, local variations in sensing and communication still benefit from personalized model aggregation.


\section{Conclusion}
\label{sec:concludes}
In this paper, we introduced an EM-based PFL framework that leverages a probabilistic aggregation mechanism to address the dual objectives of sensing and communication in ISAC systems. By the EM approach at each BS, our method adaptively balances global and local model updates in a data‐driven manner. Extensive evaluations under both homogeneous and heterogeneous settings demonstrate that the proposed approach achieves faster convergence, higher multi‐objective utility, and better robustness compared to existing baselines. In the future, we plan to extend this framework to dynamic client populations and explore its application to other multi‐task wireless scenarios. 


\bibliographystyle{IEEEtran}
\bibliography{ref}

\end{document}